\documentclass[twocolumn,aps,pra,showpacs,superscriptaddress,10pt]{revtex4-1} %

\usepackage{amsfonts}
\usepackage{subfigure}
\usepackage{amsmath}
\usepackage{txfonts}
\usepackage{amssymb}
\usepackage{amsbsy} 
\usepackage{epsfig}
\usepackage{graphicx}
\usepackage{epstopdf}
\usepackage{mathdots}
\usepackage{color}
\usepackage{cleveref}

\begin{document}

\title{Non-Hermitian Weyl Semimetals: Non-Hermitian Skin Effect and non-Bloch Bulk-Boundary Correspondence}

\author{Xiaosen Yang} \altaffiliation{yangxs@ujs.edu.cn }
\affiliation{Department of physics, Jiangsu University, Zhenjiang, 212013, China}

\author{Yang Cao}
\affiliation{Department of physics, Jiangsu University, Zhenjiang, 212013, China}

\author{Yunjia Zhai}
\affiliation{Department of physics, Jiangsu University, Zhenjiang, 212013, China}

\date{\today}

\begin{abstract}

We investigate novel features of three dimensional non-Hermitian Weyl semimetals, paying special attention to its unconventional bulk-boundary correspondence.  We use the non-Bloch Chern numbers as the tool to obtain the topological phase diagram, which is also confirmed by the energy spectra from our numerical results. It is shown that, in sharp contrast to Hermitian systems, the conventional (Bloch) bulk-boundary correspondence breaks down in non-Hermitian topological semimetals, which is caused by the non-Hermitian skin effect.  We establish the non-Bloch bulk-boundary correspondence for non-Hermitian Weyl semimetals: the Fermi-arc edge modes are determined by the non-Bloch Chern number of the bulk bands. Moreover, these Fermi-arc edge modes can manifest as the unidirectional edge motion, and their signatures are consistent with the non-Bloch bulk-boundary correspondence. Our work establish the non-Bloch bulk-boundary correspondence for non-Hermitian topological semimetals.

\end{abstract}

\maketitle

\section{Introduction}
Topological phases are characterized by bulk topological invariant. Examples are  topological insulators\cite{kanePhysRevLett2005, Bernevig1757, MoorePhysRevB2007, fuliangPhysRevLett2007, qiPhysRevB2008, HasanRevModPhys2010, qiRevModPhys2011, DzeroPhysRevLett2010, fuliangPhysRevLett2011, wangzhongPhysRevX2012}, topological superconductors/superfluids\cite{qiPhysRevLett2009, horPhysRevLett2010,SasakiPhysRevLett2011, NakosaiPhysRevLett2012, xuPhysRevLett2018, zhangfanPhysRevLett2013, zhangnaturecomm2013topological, qunaturecom2013topological}  and Weyl semimetal\cite{WanXiangangPhysRevB2011, BurkovPhysRevLett2011, GangPhysRevLett2011, xuscience2015, lvPhysRevX2015, wenhongmingPhysRevX2015, YoungPhysRevLett2015, XiaoQiPhysRevLett2015, YanPhysRevLett2016, BansilRevModPhys2016, ArmitageRevModPhys2018, YuanpingPhysRevMaterials2018, chen2018PhysRevLett}. For equilibrium closed systems, described by Hermitian Hamiltonian, the topological invariants are defined in the term of the Bloch Hamiltonian\cite{MoorePhysRevB2007,fuliangPhysRevLett2007, qiPhysRevB2008}. The Hermitian Hamiltonian has real eigenenergies and a set of orthogonal eigenstates. The bulk topological invariants dictates the existence of robust edge states at the boundary. This bulk-boundary correspondence is a ubiquitous guiding principle to the topological phases. The bulk-boundary correspondence is also applicable when the bulk is gapless, by virtue of point touching of nondegenerate conduction and valence bands\cite{WanXiangangPhysRevB2011}. The gapless bulk band structure has paired Weyl points with opposite chirality and topological charge.  The massless Weyl fermions near the Weyl points are stable against perturbations.

Recently, considerable effort has been devoted to explore the properties of nonequilibrium open systems, especially non-Hermitian systems\cite{benderPhysRevLett1998,Diehlnaturephys2011, YoungwoonPhysRevLett2010,malzardPhysRevLett2015, zhenbonature2015, huicaoRevModPhys2015,leePhysRevX2014, XuPhysRevLett2017, Caiarxiv2018,gongPhysRevX2018,ChenYuPhysRevB2018,  WeiweiPhysRevLett2018, Johanarxiv2018, CerjanPhysRevB2018,PartoPhysRevLett2018, ShunYaoarxiv2019, liuPhysRevLett2019, Kawabataurecomm2019, ZhesenPhysRevB2019}. The non-Hermitian systems include optical and mechanical systems with gain and loss\cite{MakrisPhysRevLett2008,KlaimanPhysRevLett2008, LonghiPhysRevLett2009, ruternaturephys2010, LiertzerPhysRevLett2012,Regensburgernaturephys2012, Peng328science2014,lunaturepho2014,Fleurynaturecom2015}, solid state system with finite quasiparticle lifetimes for non-Hermitian self energy\cite{Koziiarxiv2017,Papajarxiv2018,Zhouscience2018}. The non-Hermitian systems exhibit many impressive features, such as non-Hermitian skin effect\cite{ShunyuPhysRevLett2018b, ShunyuPhysRevLett2018a, MartinezAlvarez2018}, bulk Fermi arcs connecting exceptional points\cite{Koziiarxiv2017,Zhouscience2018,Kaifaarxiv2018,JohanPhysRevA2018} and biorthogonality\cite{Sokolov2006, KunstPhysRevLett2018, qiu2018, EdvardssonPhysRevB2019}. In particular, the non-Hermitian skin effect\cite{ShunyuPhysRevLett2018b, ShunyuPhysRevLett2018a, MartinezAlvarez2018} means that all energy eigenstates can be exponentially localized at the boundary of non-Hermitian systems\cite{ShenPhysRevLett2018b, WangBPhysRevB2018,haijunPhysRevB2019, Jiangbinhybrid2018, chenshuarxiv2019, Ezawaarxiv2019, Borgniaarxiv2019, chuanweiarxiv2019, yiweiarxiv2019, Ghatakarxiv2019,Murakamiarxiv2019}. The interplay between the topology and the non-Hermiticity can lead to the breakdown of the Bloch bulk-boundary correspondence\cite{ShunyuPhysRevLett2018b, ShunyuPhysRevLett2018a, MartinezAlvarez2018, JinLPhysRevB2018, KawabataPhysRevB2018, Ghatakarxiv2019, Murakamiarxiv2019}. The topological properties of the non-Hermitian systems can not be precisely predicted by the Bloch eigenstates under open boundary conditions. Furthermore, the real topological invariant is defined in non-Bloch Hamiltonian instead of Bloch Hamiltonian. The non-Bloch winding (Chern) number has been introduced to characterized the topological properties of one-dimensional (two-dimensional) systems\cite{ShunyuPhysRevLett2018b, ShunyuPhysRevLett2018a,Murakamiarxiv2019}. The non-Bloch topological invariants strictly characterize the chiral edge modes and provide the non-Bloch Bulk-boundary correspondence.

\begin{figure}[t]
\includegraphics[width=\columnwidth]{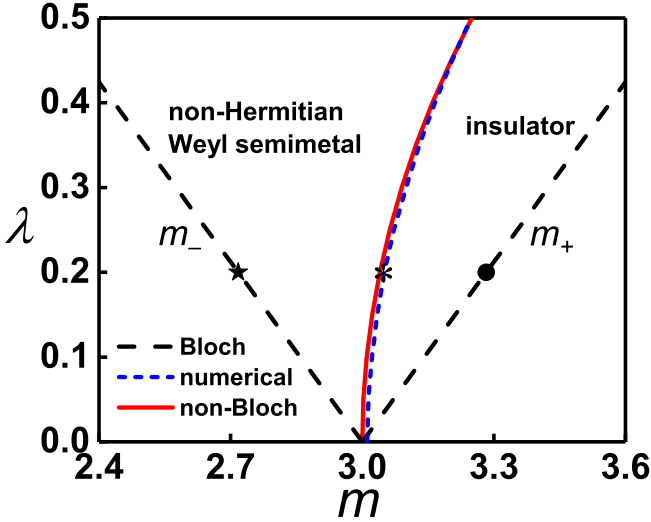}
\caption{\textbf{Topological Phase Diagram  for $\mathbf{\Lambda}  = (\lambda, \lambda, 0)$.} The blue-dotted curve, determined by the real space energy spectra of cubic open boundary, is the topological phase boundary between non-Hermitian Weyl semimetal and insulator. The left non-Hermitain Weyl semimetal has gapless bulk and gapless Fermi-arc edge modes. The right insulator has gapped bulk and gapped edge. The open-boundary
spectra for the three marked points is given in Fig.\ref{fig.3}. The topological phase boundary closely approximates to the boundary based on non-Bloch Chern number (red-solid curve with $m  = 3 + \lambda ^{2}$). This non-Bloch phase boundary is fundamentally different with the phase boundaries based on the Bloch theory (black-dashed lines with $m_{\pm} = 3 \pm \sqrt{2} \lambda$). According to Bloch band theory, the phase is non-Hermitian Weyl semimetal when $m < m_{-}$ and trivial semimetal for $m_{-} < m < m_{+}$. The phase is gapped insulator for $m > m_{+}$. The Bloch spectra have a pair of Weyl exceptional rings for non-Hermitian Weyl semimetal and a merged exceptional ring for trivial semimetal.}
\label{fig.1}
\end{figure}

For three dimensional non-Hermitian systems, the Weyl points can be spread into exceptional lines and even exceptional surfaces. Examples of non-Hermitian Weyl semimetals have been considered in Refs. \cite{XuPhysRevLett2017,CerjanPhysRevB2018,CerjanWeyl2018,Bergholtzarxiv2019}, however, their novel bulk-boundary correspondence has not been uncovered and clarified yet, which is the focus of the present paper. In this work, we investigate the topological properties of the non-Hermitian Weyl semimetal in the presence of on-site gain/loss. We analyze the shape of the exceptional rings by Bloch band theory under periodic boundary condition and give the topological phase diagram. Furthermore, the Weyl semimetal can be regarded as a stack of layers of two dimensional Chern insulator in $k_{z}$ momentum space in the absence of on-site gain/loss. Thus, we can bring insight into the topological properties of the non-Hermitian Weyl semimetal by the non-Bloch Chern number. We extend the Bloch momentum space $T^{3}(\textbf{k})$ into complex momentum space  $\tilde{T}^{3}(\tilde{\textbf{k}})$ to derive the three dimensional non-Bloch Hamiltonian. For a fixed momentum $k_{z}$, the three dimensional complex momentum space $\tilde{T}^{3}(\tilde{\textbf{k}})$ is reduced to two dimension $\tilde{T}^{2}(\tilde{\textbf{k}}_{\perp})$ in which the non-Bloch Chern number can be defined. We find a new non-Bloch bulk-boundary correspondence for the non-Hermitian Weyl semimetal. The non-Bloch Chern number can predict the Fermi-arc edge modes. As such, the Bloch band theory and the conventional bulk-boundary correspondence breaks down for the non-Hermitian skin effect, which fundamentally affects the topological phase diagram. The validity of non-Bloch Chern number is confirmed by comparing its prediction to numerical results of real space energy spectra and edge-state transport.

\section{Non-Hermitian Bloch Hamiltonian}

We consider a non-Hermitian Bloch Hamiltonian of semimetal on a cubic lattice:
\begin{eqnarray}
H(\textbf{k})&=& (\sin k_{x} + i \lambda)\sigma_{x} + (\sin k_{y} + i \lambda) \sigma_{y}\nonumber\\
& &+(m - \cos k_{x} - \cos k_{y} - \cos k_{z})\sigma_{z},
\label{hamiltonian}
\end{eqnarray}
where $\sigma_{x,y,z}$  are Pauli matrices. The non-Hermitian parameters $\mathbf{\Lambda} = (\lambda, \lambda, 0)$ appear as 'imaginary Zeeman fields' strength. In the absence of non-Hermitian parts ($\mathbf{\Lambda} = 0$), the eigenvalues of the system are $E_{\pm}(\textbf{k})=\pm h (\textbf{k})$ with $h(\textbf{k}) = |\textbf{h(\textbf{k})}|$ and $\textbf{h(\textbf{k})} = (\sin k_{x}, \sin k_{y}, m - \cos k_{x} - \cos k_{y} - \cos k_{z})$. A pair of Weyl points with $Z_{2}$ topological charge are stable when $|m| < 3$. The topological nontrivial phase is Weyl semimetal. When $|m| > 3$, the two Weyl points will annihilate with each other and the phase is gapped insulator. There is a topological phase transition between Weyl semimetal and insulator at $|m| = 3$. Therefore, we will focus on $m$ being close to $3$.

\begin{figure}[t]
\includegraphics[width=\columnwidth]{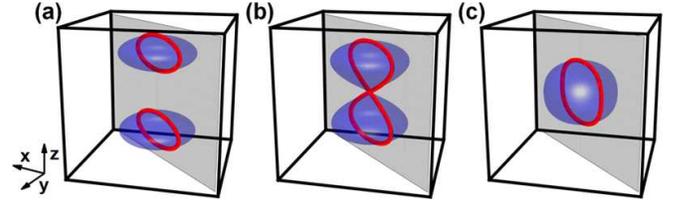}
\caption{Illustration of the exceptional rings in Bloch Brillouin zone with $\lambda = 0.2$ for (a) $m = 2.6$, (b) $m = 2.7172$ and (c) $m = 3.05$. In the presence of the non-Hermitian part, the paired Weyl points are spread into a pair of Weyl exceptional rings with opposite charge. As increasing of $m$, the two Weyl exceptional rings will merge into one uncharged exceptional ring at $m = m_{-}$. The uncharged exceptional ring will disappear when $m > m_{+}$.}  \label{fig.2}
\end{figure}

In the presence of non-Hermitian parts, the Bloch energies of above $H(\textbf{k})$ are
$E_{\pm}(\textbf{k}) = \pm \sqrt{h(\textbf{k})^{2} - \Lambda^{2} + 2 i \mathbf{\Lambda} \cdot \textbf{h(\textbf{k})}}$.
A non-Hermitian band is called  "fully gapped" (or "isolated") if the energy has no overlap with that of any other bands in the complex-energy plane, while is called "gapless" (or "inseparable") if the complex-energy is degenerate with other bands\cite{ShenPhysRevLett2018b}.
For our non-Hermitian system, the Bloch bands are gapless when $m < m_{+}$ with $m_{\pm} = 3 \pm \sqrt{2} \lambda$ as shown in Fig.~\ref{fig.1}. For the gapless regions, the phase is non-Hermitian Weyl semimetal and the Bloch spectra have a pair of Weyl exceptional rings with opposite charge\cite{CerjanPhysRevB2018} for periodic boundary condition when $m < m_{-}$. The two Weyl exceptional rings merge into one uncharged exceptional ring and the phase is topological trivial semimetal for $m_{-} < m < m_{+}$. There is a topological phase boundary between non-Hermitian Weyl semimetal and trivial semimetal at $m = m_{-}$.  The uncharged exceptional ring will shrink into an exceptional point at $m = m_{+}$. Then, the phase is gapped insulator when $m > m_{+}$. Fig.~\ref{fig.2} shows the evolution of the exceptional rings as the increasing of $m$ with $\lambda = 0.2$.

\section{Phase diagram based on non-Bloch Chern number}

In Hermitian systems, Weyl semimetals are characterized by topologically protected Fermi-arcs. The chiral/helical gapless edge states exist in a finite region in momentum space and should be determined by the properties of Bloch Hamiltonian. The Bloch Bulk-boundary correspondence is a key properties of Weyl semimetals.  However, the Bloch bulk-boundary correspondence is not applicable to the topological properties of the non-Hermitian system for non-Hermitian skin-effect. Therefore, the topological phase diagram based on the Bloch band theory will generate pronounced deviation to the real phase diagram. Thus, we draw the phase boundaries by the non-Bloch Chern number and the real space energy spectra.

The non-Bloch Chern number is defined in a complex momentum space instead of Bloch momentum space\cite{ShunyuPhysRevLett2018a}. To determine the topological phase boundary, we consider the low-energy continuum case in $x-y$ plane of our non-Hermitian Bloch Hamiltonian Eq.(\ref{hamiltonian}), which can be rewritten as following:
\begin{eqnarray}
H(\textbf{k})&=& (k_{x} + i \lambda)\sigma_{x} + (k_{y} + i \lambda) \sigma_{y}\nonumber\\
& &+\left( m - 2 + \frac{k_{x}^{2}}{2} + \frac{k_{y}^{2}}{2} - \cos k_{z} \right)\sigma_{z},
\label{hamiltonian1}
\end{eqnarray}

To extend the Bloch momentum space into complex momentum space, we take $k_{x,y} \rightarrow \tilde{k}_{x,y} + i \tilde{k}'_{x,y}$ and $k_{z} \rightarrow \tilde{k}_{z}$. Here, the imaginary parts take the form $\tilde{k}'_{x,y} = - \lambda$\cite{ShunyuPhysRevLett2018a}. The Bloch Brillouin zone $T^{3}(\textbf{k})$ undergoes a deformation to non-Bloch Brillouin zone $\tilde{T}^{3}(\tilde{\textbf{k}})$. The non-Bloch Hamiltonian is defined as follows:

\begin{eqnarray}
\tilde{H}(\tilde{\textbf{k}}) &\equiv& H(\textbf{k} \rightarrow \tilde{\textbf{k}} + i \tilde{\textbf{k}'}),\nonumber\\
 &=& \tilde{k}_{x} \sigma_{x} + \tilde{k}_{y} \sigma_{y}\nonumber + \left[\tilde{m}(\tilde{k}_{z}) + \frac{\tilde{k}_{x}^{2} + \tilde{k}_{y}^{2}}{2} - i \lambda (\tilde{k}_{x} + \tilde{k}_{y})\right]\sigma_{z},
\label{hamiltonian2}
\end{eqnarray}
with $\tilde{m}(\tilde{k}_{z}) = m - 2 - \lambda^{2} - \cos \tilde{k}_{z}$. The right/left eigenvectors of the non-Hermitian Hamiltonian are $\left|u_{R n}(\tilde{\textbf{k}})\right>$ $\Big/$ $\left<u_{L n}(\tilde{\textbf{k}})\right|$ and satisfy:
\begin{eqnarray}
\tilde{H}(\tilde{\textbf{k}}) \left|u_{R n}(\tilde{\textbf{k}})\right> = E_{n} \left|u_{R n}(\tilde{\textbf{k}})\right>, \nonumber\\ \tilde{H}^{\dag}(\tilde{\textbf{k}}) \left|u_{L n}(\tilde{\textbf{k}})\right> = E_{n}^{*} \left|u_{L n}(\tilde{\textbf{k}})\right>,\nonumber
\label{H2}
\end{eqnarray}
where $n$ is the band index. The eigenvectors are normalized with $\left<u_{L n}|u_{R m}\right> = \delta_{nm}$. We have $\tilde{k}_{z} = k_{z}$, which is real.  For a fixed real value of $k_{z}$, the non-Bloch Brillouin zone $\tilde{T}^{3}(\tilde{\textbf{k}})$ is reduced to $\tilde{T}^{2}(\tilde{\textbf{k}}_{\perp})$  with $\tilde{\textbf{k}}_{\perp} = (\tilde{k}_{x} ,\tilde{k}_{y})$. Then, we use the definition of non-Bloch Chern number in two non-Bloch Brillouin zone $\tilde{T}^{2}(\tilde{\textbf{k}}_{\perp})$ \cite{ShunyuPhysRevLett2018a}:
\begin{eqnarray}
C_{n}(k_{z}) = \frac{1}{2\pi i} \int _{\tilde{T}^{2}} d^{2} \tilde{\textbf{k}}_{\perp} \epsilon^{i,j} \left<\partial_{i} u_{L n}(\tilde{\textbf{k}}_{\perp}, k_{z}) \big| \partial_{j} u_{R n} (\tilde{\textbf{k}}_{\perp}, k_{z})\right>,
\label{nonblochchernnumber}
\end{eqnarray}
where $\epsilon^{x,y} = - \epsilon^{y,x} = 1$. For a fixed $k_{z}$, the non-Bloch band is "fully gapped". The non-Bloch Chern number is $0$ for $\tilde{m}(k_{z}) > 0$ and $1$ for $\tilde{m}(k_{z}) < 0$. As $k_{z}$ is varied, $C(k_{z})$ remains constant as long as the gap is unclosed. $C(k_{z})$ will change only when the gap is closed at $k_{z} = \pm k_{z}^{c}$ and the $\tilde{T}^{2}(\tilde{\textbf{k}}_{\perp})$ plane crosses non-Hermitian Weyl nodes. Like the case of Hermitian Weyl semimetal, we also can assign to the non-Hermitian Weyl nodes an integer topological charge (equal to the change of the non-Bloch Chern number across the Weyl nodes). The topological charge stabilizes the non-Hermitian Weyl nodes. Therefore, the topological phase boundary between non-Hermitian Weyl semimetal and insulator based on non-Bloch Chern number is
\begin{eqnarray}
m = 3 + \lambda^{2},
\label{nonblochB}
\end{eqnarray}
which is shown in Fig.~\ref{fig.1}. The non-Hermitain Weyl semimetal phase on the left of the topological boundary has gapless bulk and gapless Fermi-arc edge modes. The insulator phase on the right has gapped bulk and gapped edge. There is a dichotomy between the two topological phase diagrams bases on the non-Bloch Chern number and the Bloch band theory. The exact topological phase boundary is only a single curve and the phase diagram has no topological trivial semimetal. Interesting, tn sharp contrast to the Hermitian systems, the conventional bulk-boundary correspondence breaks down in the non-Hermitian Weyl semimetal. The Fermi-arc edge modes of the non-Hermitian Weyl semimetal are determined by the non-Bloch Chern number of the bulk bands. The breakdown of the Bloch band theory is caused by the non-Hermitian skin effect.

\begin{figure}[t]
\includegraphics[width=\columnwidth]{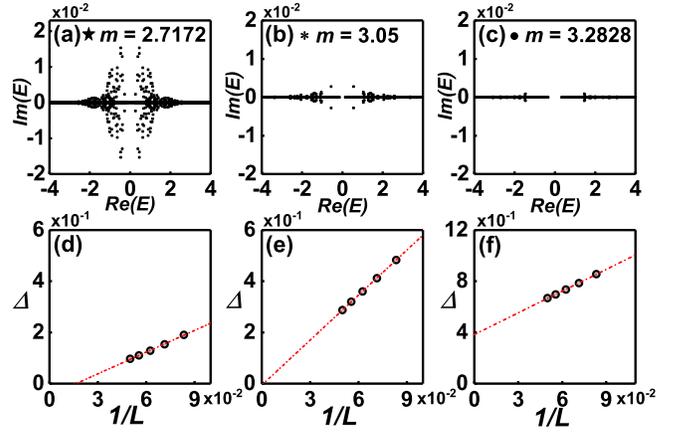}
\caption{\textbf{Real Space Energy Spectra Under Cubic Open-Boundary Condition: } (a) $m=2.7172$, (b) $m=3.05$ and (c) $m=3.2828$ (three value of parameters indicated in Fig.~\ref{fig.1})  with $\lambda = 0.2$ and lattice site size $L=20$. (d) - (f) show the magnitude of the gap $\Delta$ as functions of $1/L$ for (a) - (c), respectively. $L$ is the lattice site size in $x,y,z$ direction. The intercept of $\Delta-1/L$ line gives the gap in large scale limit ($L \rightarrow \infty$). The gap is zero for (d)(e) and nonzero for (f).}  \label{fig.3}
\end{figure}

\begin{figure*}[t]
\includegraphics[width=\textwidth]{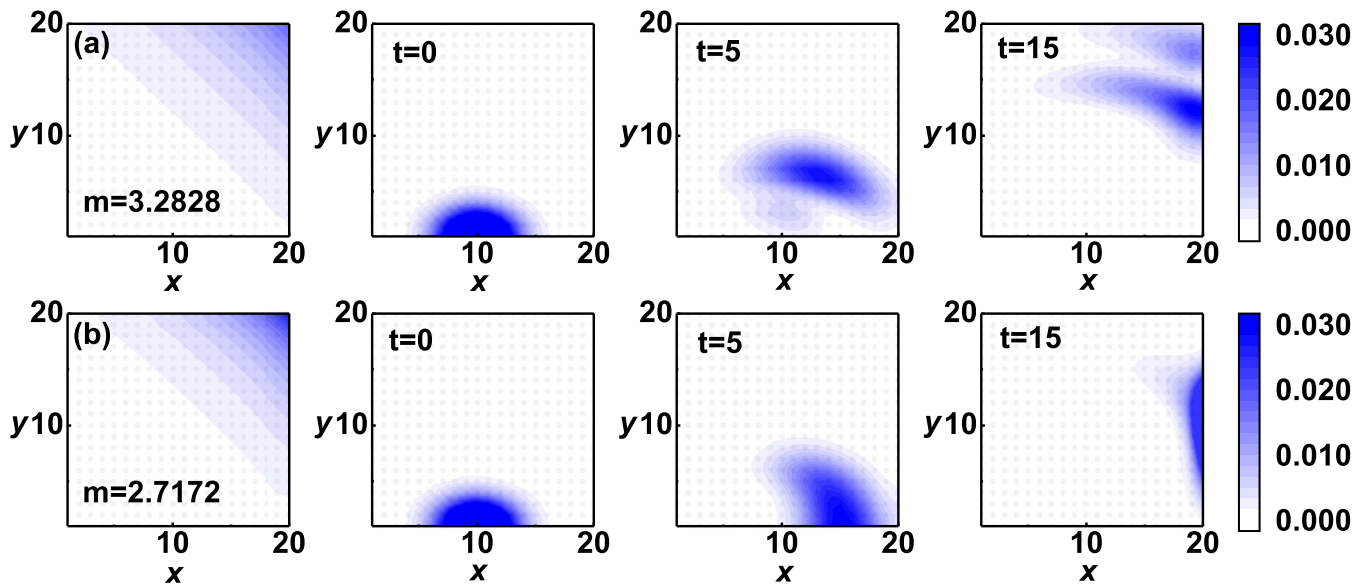}
\caption{\textbf{Non-Hermitian Skin Effects.} Left panel: total normalized eigenstates $ \mathcal{N}_{0} \sum_{n} \left| u_{R n} \right>$ under cubic open-boundary condition for (a) $m = 3.2828$ and (b) $m = 2.7172$ with $\lambda=0.2$ and lattice sites size $L = 20$. Right panels: wave pocket evolutions. The initial wave pocket takes the Gaussian form $\psi(t=0)=\mathcal{N}\exp [-(i_{x}-10)/20 -(i_{y}-1)/10 -(i_{z}-10)/20](1,1)^{T}$, normalized by $\mathcal{N}$. The modulus squared intensity of $\left|\psi(t)\right>$ is normalized and shown for $t=0, 5, 15$ in $x-y$ plane by adding up in $z-$direction. The wave packet fades into the bulk in up row. There has unidirectional edge motion in down row.} \label{fig.4}
\end{figure*}

To check the valid of topological phase diagram based on non-Bloch Chern number, we calculate the real space energy spectra under cubic open-boundary condition (lattice sites size $L \times L \times L$). The up row of Fig.~\ref{fig.3} show the spectra for (a) $m = 2.7172$, (b) $m = 3.05$ and (c) $m = 3.2828$ with $\lambda = 0.2$ (three indicated points in Fig.~\ref{fig.1}). Considering the size effects, we make the lattice size scaling of the gap in Fig.~\ref{fig.3}(d) - (f) for (a) - (c), respectively. The gap is given by the intercept of $\Delta - 1/L$ line. In Bloch theory, the three spectra are gapless and have exceptional rings/point in the spectra.  As shown in Fig.~\ref{fig.3}(a) and (b), the gap vanishes for $m = 2.7172$ and $m = 3.05$ when $L \rightarrow \infty$. Remarkably, there is a clear gap at the spectra of $m = 3.2828$. Base on the spectra of cubic open-boundary condition, there is a topological phase transition between the gapless non-Hermitian Weyl semimetal and gapped insulator phases at $m = 3.05$. We draw the gapless-gaped phase boundary under cubic open-boundary condition by blue-dotted curve in Fig.~\ref{fig.1}. The two curves base on the open-boundary energy spectra and the non-Bloch Chern number are very close. Therefore, the non-Bloch Chern number is valid to our three dimensional non-Hermitian Weyl semimetal.

\section{Non-Hermitian skin effect and transport signature of edge modes}

Different from the Hermitian Hamiltonian, the eigenstates are non-orthogonal for the non-Hermitian cases. All the eigenstates are exponentially localized at the boundary of the system. To illustrate the non-Hermitian skin effect, Fig.~\ref{fig.4} shows the bulk states in $x-y$ plane by adding up $z-$direction for (a) $m = 3.2828$ and (b) $m = 2.7172$ with $\lambda=0.2$.  The bulk states are localized at the boundary for both the non-Hermitian Weyl semimental and topological trivial gapped insulator phases. The usual bulk-boundary correspondence is invalid for non-Hermitian systems.

For topological nontrivial phase, the localized eigenstates have Fermi-arc edge modes and gapless bulk states.  However, there is no Fermi-arc edge modes in the topologically trivial regime. The chirality of the Fermi-arc edge modes will affect the wave pocket time evolution in topological nontrivial phases. To reveal the topological properties and the Fermi-arc edge modes, we investigate the wave pocket time evolution. The time dependent wave satisfies the non-Hermitian Schr$\ddot{o}$dinger equation:
\begin{eqnarray}
i \partial_{t}\left|\psi(t)\right> = H \left|\psi(t)\right>.
\label{Schrodinger}
\end{eqnarray}

For an initial wave $\left|\psi(t=0)\right>$, the time dependent wave function is $\left|\psi(t)\right> = \sum_{n} \exp(- i E_{n} t)\left|u_{R n}\right> \left< u_{L n} | \psi(t=0)\right>$. As shown in Fig.~\ref{fig.4}, the wave pocket quickly spread into the bulk for topological trivial insulator phase with $m=3.2828$. However, there is clear chiral edge motion for the topological non-Hermitian Weyl semimetal with $m=2.7172$. This can be explain as following. Despite the eigenstates are localized at the boundary for topological trivially phase, there is no signature of chiral edge motion.  Thus, the wave pocket evolve to the bulk states by quick enter into the bulk without any topological constrain. For topological nontrivial phase, there is chiral edge modes with zero energy. The chirality of the edge modes will constrain the wave pocket evolution along the edge. The existence/absence of the chiral edge motion can be used to determine the non-Hermitian topological nontrivial phases in theory and future experiment.

\section{Conclusion}

We investigated the novel features of three dimensional non-Hermitian Weyl semimetals by non-Bloch Chern number, Bloch band theory, open-boundary energy spectra and dynamics. We showed that the non-Hermitain Weyl semimetals have gapless bulk and gapless Fermi-arc edge modes. We uncover the non-Bloch bulk-boundary correspondence for the non-Hermitian Weyl semimetal. The Fermi-arc edge modes of the non-Hermitian Weyl semimetal are strictly determined by the non-Bloch Chern number of the bulk bands. Thus, the conventional bulk-boundary correspondence breaks down for the non-Hermitian skin effect.  The non-Hermitian skin effect also generates pronounced deviation of the phase diagram from the Bloch band theory. The topological phase transition between nontrivial and trivial phases does not occur at the two Bloch phase boundaries. The topological phase boundary is only a single curve in the phase diagram. The valid of the non-Bloch Chern number is confirmed by the cubic open-boundary energy spectra. Furthermore, we showed that the Fermi-arc edge modes can manifest as the unidirectional edge motion.

\section*{Acknowledgment}
We would like to thank Zhong Wang for fruitful discussion. This work is supported by NSFC under Grants No.11504143.

\bibliographystyle{apsrev4-1}
\bibliography{reference}

\end{document}